\documentclass[11pt]{article}
\usepackage{amstex,amssymb,cite,epsf}

\numberwithin{equation}{section}

\begin{document}
\begin{flushright}
IMSc/2000/02/03 \\
TIFR/TH/00-08 \\
\end{flushright}
\begin{center}
{\large{\bf HORIZON STATES FOR AdS BLACK HOLES}}

\bigskip 
T. R. Govindarajan$^a$\footnote{trg@@imsc.ernet.in},
V.Suneeta$^a$\footnote{suneeta@@imsc.ernet.in} and 
S. Vaidya$^b$\footnote{sachin@@theory.tifr.res.in}\\ 

$^a${\it Institute of Mathematical Sciences, \\
Chennai, 600 113, India.}

$^b${\it Tata Institute of Fundamental Research, \\
Colaba, Mumbai, 400 005, India.}
\end{center}

\begin{abstract}
We study the time-independent modes of a massless scalar field in
various black hole backgrounds, and show that for these black holes,
the time-independent mode is localized at the horizon. A similar
analysis is done for time-independent, equilibrium modes of the
five-dimensional plane AdS black hole. A self-adjointness analysis of
this problem reveals that in addition to the modes corresponding to
the usual glueball states, there is a discrete infinity of other
equilibrium modes with imaginary mass for the glueball. We suggest
these modes may be related to a Savvidy-Nielsen-Olesen-like 
vacuum
instability in QCD.
\end{abstract} 

\section{Introduction}
A study of various kinds of matter fields propagating in black hole
backgrounds yields information about diverse classical and quantum
aspects of black hole physics. Detailed analysis of modes of the
scalar, spinor and gauge fields in black hole backgrounds can be found
for example, in \cite{chandra}. In particular, for scalar fields, the
energies of these modes are given by the square root of the
eigenvalues of the spatial part of the Klein-Gordon operator in that
background. For static spacetimes with null singularities, it has been
argued \cite{wald,horo} that the spatial part of the Klein-Gordon
operator is essentially self-adjoint. Further, since it is positive
and symmetric, one can choose a positive self-adjoint extension such
that the eigenvalues are all positive and hence the energies
real. However, as we show in this paper, in the case of many black
hole spacetimes, near the null singularity at the horizon, the zero
(time-independent) mode of the scalar field has to be handled
separately. In particular, the boundary conditions imposed on the zero
mode both at the horizon and at infinity are different from those on
the other modes with real energies. In fact, we will show that there
are an infinite number of boundary conditions, labeled by a $U(1)$
parameter, that lead to one zero mode solution. This solution could be
thought of as a `horizon state' as it is localized at the horizon.

An application of this analysis to the time-independent solutions of
the infinite mass limit of the AdS-Schwarzschild
black hole \cite{hawking} leads to results that could be
interesting in light of the AdS/CFT correspondence. As proposed by
Witten \cite{witten}, the AdS/CFT duality relating supergravity on
anti-de Sitter space to a supersymmetric Yang-Mills theory on the
boundary can be extended to non-supersymmetric $QCD$. The AdS
background is replaced by an AdS-Schwarzschild black hole background.
It has been shown that gravity on this background gives many of the
features of strong coupling limit of $QCD$, like the area law behavior
of Wilson loops, confinement, and the glueball mass spectrum with a
mass gap \cite{witten,ooguri,jevicki,mathur}.

The glueball mass spectrum is reproduced by certain time-independent
and normalizable modes obtained by solving the dilaton wave equation
in the black hole geometry. These modes were numerically computed
first in \cite{ooguri,jevicki}. These modes are ``equilibrium modes''
for the black hole, i.e. the current vanishes at the horizon, which
has been recognised in \cite{mathur} as the correct boundary condition
to be used at the horizon.

It has been argued that ``non-equilibrium'' modes of the same black
hole (with ingoing boundary conditions at the horizon) give the time
scale of approach to thermal equilibrium of the boundary Yang-Mills
theory. These modes, i.e. the quasi-normal modes of the black hole,
have been computed recently \cite{bala,hubeny}.

In this paper, we study the scalar wave equation in the
AdS-Schwarzschild background, and show, that written as a Hamiltonian
problem, it is not self-adjoint. Self-adjointness and completeness
requires inclusion of modes ignored in
\cite{ooguri,jevicki,mathur}. These modes are also equilibrium modes
of the black hole but are irregular at the horizon $^{}$\footnote{In
\cite{jevicki,mathur}, the existence of irregular modes is
mentioned. However, they are not considered.}.  They are also
tachyonic.  We suggest that these modes are dual in the AdS/CFT sense
to modes in $QCD_3$ signaling the onset of a
Savvidy-Nielsen-Olesen-like instability of the vacuum
\cite{savvidy,nn,no,ano}.

The organization of the paper is as follows. In section 2, we briefly
describe two kinds of modes that are commonly discussed in related
literature, namely the normalizable equilibrium modes, and the
non-normalizable quasi-normal modes, to emphasize the differences
between them. We also show that for a massless scalar field
propagating in Schwarzschild or Reissner-Nordstrom black hole
background, the Klein-Gordon operator is self-adjoint. In section 3,
we focus on the zero energy mode of the scalar field in these
backgrounds, and in the background of the (1+1)-d black hole
\cite{witten3} as well as the BTZ black hole \cite{btz}. The equation
obeyed by the zero mode has unusual properties, which we analyze in
section 4. In particular, we show that this state localized at the
horizon. In section 5, we apply the results of section 4 to study the
zero mode of the massless scalar field in the background of the
infinite mass limit of the AdS-Schwarzschild black hole, and argue
that the ``horizon states'' are necessary for completeness. In section
7, we speculate on the interpretation of these irregular modes in the
boundary theory, and suggest that they may be related to a
Savvidy-Nielsen-Olesen-like instability.

\section{Modes of the scalar field in black hole background}
As mentioned before, the energies of {\em normalizable} modes of a
scalar field in the exterior of a black hole spacetime (i.e. in the
region from the outer horizon to infinity) have real energies.

This can be verified for the Schwarzschild or Reissner-Nordstrom 
black hole in the exterior. There are no normalizable mode solutions
with complex (or pure imaginary) energies. However, this is not true
in a region of the black-hole spacetime near a timelike singularity.
For the Reissner-Nordstrom spacetime, in the region between the
timelike singularity and the inner horizon, the spatial part of the
Klein-Gordon operator is not self-adjoint, as also observed by
\cite{jacobson}, but can be made self-adjoint by a suitable choice of
boundary conditions. There exist boundary conditions for which there
is a negative eigenvalue for this operator, leading to a mode solution
with imaginary energy. However, this solution is not extendible to the
physical region of interest between the outer horizon and infinity.

Other modes of importance in the context of black holes are the
quasi-normal modes (see for example, \cite{kokkotas}). For the case of
asymptotically flat black holes, these are defined to be purely
ingoing near the horizon and outgoing at infinity. These are not
normalizable, but are of interest as their energies are the
characteristic frequencies associated with the perturbation of the
black hole. These are in general, complex, and decay with time. In the
Schwarzschild and Reissner-Nordstrom cases, there are an infinite
number of such modes (see \cite{kokkotas} for references) which
include purely imaginary modes \cite{chandra1}.

Recently, quasi-normal modes for the AdS-Schwarzschild black hole have
also been studied \cite{hubeny}. These are different from the
quasi-normal modes for asymptotically flat black holes, in that they
are not outgoing at infinity, but vanish. This is due to the fact that
the AdS potential diverges at infinity. Numerical results of
\cite{hubeny} suggest that these modes are complex.  However, they are
still non-normalizable due to their behavior at the horizon.

An analysis of the spatial part of the Klein-Gordon operator for the
AdS-Schwarzschild black hole shows that as expected in \cite{horo},
the operator is self-adjoint, and all the normalizable modes have real
energies. The AdS-Schwarzschild black hole has a metric
\begin{eqnarray}
ds^2 &=& - F(r)d\tau^2 + F^{-1}(r) dr^2 + r^2 d\Omega^2, \quad {\rm
where} \\ \label{adsbh1}
F(r) &=& (1 + r^2/b^2 - r_{0}^2/r^2).
\end{eqnarray}	
Here $b$ is the radius of curvature of the anti-de Sitter space and
$r_{0}$ is related to the black hole mass,
\begin{equation}	
M = \frac{3 A_{3} r_{0}^2}{16 \pi G_{5}}
\end{equation} 
and $A_{3}$ is the area of a unit 3-sphere.

Let us look at a massless scalar field in this background geometry.
One can in principle consider a complex scalar field with charge $q$
and mass $m$, but for simplicity we shall consider only the massless
and uncharged field in the black hole background. The action for such
a field $\Phi$ is
\begin{eqnarray} 
S &=& -\frac{1}{2} \int \sqrt{|g|} g^{ij} (\partial_i \Phi) (\partial_j
\Phi) d^5 x,\\ 
&=& -\frac{1}{2} \int_{r_{+}}^{\infty}dr \int dt \int d\Omega
\left[ r^3 \left\{ \frac{\dot{\Phi}^2}{F} + F {\Phi'}^2 + (1/r^2) \Phi
L^2 \Phi \right\}  \right]. \nonumber 
\end{eqnarray} 

The Klein-Gordon equation for the field $\Phi$ can be obtained from
above. On making the ansatz
\begin{equation}
\Phi=\frac{f(r)}{r^{3/2}}~ Y(angles)~\exp -(i \omega t),
\end{equation}
the wave functions are defined on the measure $dr/F$. The Klein-Gordon
equation can then be written in terms of the tortoise coordinate
$r_{*}$, which is defined by $dr_{*}~=~dr/H$. It takes the form
\begin{equation}
- \frac{d^2}{dr_{*}^2} f + V(r_{*}) f~~ =~~ \omega^2 f,
\label{kg1}
\end{equation}
with the measure now being $dr_{*}$. The potential is positive,
vanishes at the horizon $r_{*}= - \infty$ and diverges at $r =
\infty$. This corresponds to a finite $r_{*}$ and therefore the
solutions have to vanish there. Multiplying (\ref{kg1}) by the complex
conjugate of $f$ and integrating over the spacetime from the horizon
to infinity, it can be seen that there can be no normalizable
solutions that correspond to $\omega^2$ negative or complex. This is
in conformity with the fact that the Klein-Gordon operator is
self-adjoint.

\section{Time independent mode in black hole solutions}
The positive energy solutions to the Klein-Gordon equation can be
analyzed for most black hole solutions by going to the tortoise
coordinate $r_{*}$ mentioned in the previous section. The solutions
behave as $f \sim \exp i\omega (t \pm r_{*})$ near the horizon and
near infinity.  The horizon is at $r_{*} = - \infty$, while the
infinity of the Schwarzschild radial coordinate is either at $r_{*} =
\infty$ or at a finite $r_{*}$, depending on the black hole
considered. The solutions are plane wave normalizable, and have
infinitely oscillating phases at the horizon.

The near-horizon analysis of black hole solutions reveals, however,
that the time independent ($\omega = 0$) mode of the scalar field has
to be handled carefully.

The metric for an asymptotically flat, spherically symmetric, static
black hole in 4-D is of the form
\begin{equation}
ds^2 = -F(r)dt^2 + F^{-1}(r) dr^2 + r^2 d \Omega^2 \equiv g_{ij}
dx^i dx^j.
\end{equation}
For a Reissner-Nordstrom black hole,
\begin{eqnarray}
F(r) &=& -\frac{(r-r_{+})(r - r_{-})}{r^2}, \\
r_{\pm} &=& Q l_P + E l_P \pm (2QEl^3_P + E^2 l^4_P)^2.
\end{eqnarray}
Here, $l_P$ is the Planck length and $E= M-Q/l_P$ is the energy above
extremality. For a Schwarzschild black hole, $F(r) = (1 - 2M/r)$.

Let us look at a massless scalar field in this background geometry.
The action for such a field $\phi$ is
\begin{equation}
S= -\frac{1}{2} \int \sqrt{|g|} g^{ij} \partial_i \phi  \partial_j \phi.
\end{equation}

If we restrict out attention to spherically symmetric configurations,
the action looks like
\begin{equation}
S= -\frac{1}{2} \int \left[ -(\dot{\phi})^2 + F^2(r) (\phi')^2 \right]
\frac{dr}{F(r)} dt.
\end{equation}
This immediately allows us to identify the Lagrangian: 
\begin{equation} 
L = \frac{1}{2} \int \frac{dr}{F(r)}\left[(\dot{\phi})^2 - F(r)^2
(\phi')^2 \right].
\end{equation} 
The modes of the scalar field are obtained from the ansatz that the
time dependence of $\phi$ is $\phi \sim \exp (i\omega t)$. We are
interested in the time-independent solutions, so we take $\omega = 0$.
Then the Klein-Gordon equation for this case is obtained simply by
considering the second term in the Lagrangian, and is
\begin{equation}
H = -\frac{1}{F} \frac{d}{dr} \left( F \frac{d}{dr} \right) \psi = 0,
\end{equation}
where wave functions are defined on $L^2[(0, \infty), r^2 F dr]$. It
is more convenient to work with the measure $dr$ rather than $r^2
F(r)dr$, so we make a unitary transformation from $L^2[\mathbb{R}^+ ,
F(r)dr]$ to $L^2[\mathbb{R}^+, dr]$ via $U\psi = \sqrt{r^2 F(r)} \psi
= \chi$. In this new basis, $H$ reads:
\begin{eqnarray}
H = -\frac{d^2 \chi}{dr^2} + \left[ \frac{(r^2 F)''}{2F} - (\frac{(r^2
F)'}{2F})^2 \right] \chi = 0.
\label{Feqn}
\end{eqnarray}

On putting the value of $F$ for the black hole in (\ref{Feqn}) and
taking the near-horizon limit, we find that both for the non-extremal
black holes, (\ref{Feqn}) in the near-horizon limit is
\begin{equation}
\left(-\frac{d^2}{dx^2}  - \frac{1}{4x^2} \right) \chi = 0,
\end{equation}
where $x=(r - r_{+})$ is the near-horizon coordinate. $r_{+}$ is the
 horizon. For the extremal Reissner-Nordstrom solution, however,
 (\ref{Feqn}) reduces near the horizon to
\begin{equation}	
 - \frac{d^2 \chi}{dx^2} = 0.
\label{spham}
\end{equation} 

Another situation where we see a similar equation is the near horizon
geometry of the one-dimensional black hole discovered by Witten
\cite{witten3}. The metric for this black hole is of the form
\begin{equation}
ds^2 = - \tanh^2(r/R)dt^2 + dr^2.
\end{equation}
The action for a scalar field propagating in this background is 
\begin{equation}
S = -1/2 \int \sqrt{|g|}g^{ij}\partial_i\phi \partial_j \phi dr dt.
\end{equation} 
The Lagrangian is 
\begin{equation}
L = 1/2 \int \tanh(r/R)\left[ \frac{\dot{\phi}^2}{\tanh^2(r/R)} -
{\phi'}^2 \right] dr.
\end{equation}
The Klein-Gordon equation for the zero mode can be calculated from the
2nd term, the functions being defined on $L^2[(0, \infty),
\tanh(r/R)dr]$:
\begin{equation}
-\frac{1}{\tanh(r/R)}\frac{d }{dr} \left[ \tanh(r/R)\frac{d }{dr} 
\right] \psi = 0. 
\end{equation}

Again, we can make a unitary transformation from $L^2[\mathbb{R}^+,
\tanh(r/R)dr]$ to $L^2[\mathbb{R}^+, dr]$ via $U \psi =
\sqrt{\tanh(r/R)}\psi = \chi$, the equation now is
\begin{equation}
-\frac{d^2 \chi}{dr^2} + \frac{1}{R^2} \left[ \frac{-1/4}{\tanh^2(r/R)} +
\frac{3}{4}\tanh^2(r/R) - \frac{1}{2} \right] \chi = 0.
\end{equation} 
For small $r$, the equation is approximately
\begin{equation}
- \frac{d^2 \chi}{dr^2} - \left[\frac{1}{4 r^2} + \frac{1}{2R^2}
\right]\chi = 0.
\end{equation}

Another black hole that exhibits the same behavior is the BTZ black
hole in $(2+1)-D$ gravity \cite{btz}. For simplicity, we take $J = 0$.
It has a metric given by
\begin{equation} 
ds^2 = - N^2 dt^2 + 1/N^2 dr^2 + r^2 d\phi^2,
\end{equation} 
where $N^2 = (r^2/l^2 - M)$, $-1/l^2$ is the curvature of AdS space
and $M$ is the black hole mass. Here, again, the near horizon Klein-
Gordon equation is
\begin{equation} 
- \frac{d^2 \psi}{dx^2} - \frac{\psi}{4x^2} = 0,
\label{conformal}
\end{equation} 
where $(r - l \sqrt M) = x$ is the near-horizon coordinate.  In the
case of the Schwarzschild, non-extremal and Reissner-Nordstrom
equations, the next-order correction is of order $1/(r - r_{+})$.

Thus, in all these cases barring the extremal RN black hole,
(\ref{conformal}) is the near-horizon equation for the zero-mode
solution. The solutions, both to (\ref{conformal}) and the extremal
case are discussed in the next section. The eigenvalues of the
Hamiltonian which is just the l.h.s of (\ref{conformal}) and of the
operator which is the l.h.s of (\ref{spham}) are obtained. The
solutions of interest are the zero eigenvalue solutions for that
Hamiltonian problem. We will see that the self-adjointness analysis of
the Hamiltonian $H$, i.e the l.h.s operator in (\ref{conformal}) will
help us find these solutions.

\section{Self-Adjointness of the operator $H$}
As is well-known (see for example \cite{RS1,RS2}), discussion of
self-adjointness (or ``hermiticity'') for an unbounded operator ${\cal
O}$ first requires us to define the domain $D({\cal O})$ of ${\cal
O}$. We will only be interested in operators that are defined on
domains that are dense in the Hilbert space. This allows us to define
${\cal O}^*$, the adjoint of ${\cal O}$, and $D({\cal O}^*)$. By
definition, ${\cal O}$ is self-adjoint if and only if $D({\cal O}) =
D({\cal O}^*)$. A better way of saying this is by looking at
``deficiency indices'', which are defined as follows. Let ${\cal
K}_{\pm}=Ker (i \pm {\cal O}^*)$, where $Ker (X)$ is the kernel of the
operator $X$. The integers $n_{\pm} \equiv dim \;{\cal K}_{\pm}$ are
the deficiency indices of the operator. If $n_{\pm}=0$, then ${\cal
O}$ is essentially self-adjoint. If $n_{+} =n_{-}=n \neq 0$, the
${\cal O}$ is not self-adjoint but has self-adjoint
extensions. Different self-adjoint extensions of the operator are in
one-one correspondence with unitary maps from ${\cal K}_{+}$ to ${\cal
K}_{-}$, that is, they are labeled by a $U(n)$ matrix. Finally, if
$n_{+} \neq n_{-}$, then ${\cal O}$ cannot be made self-adjoint.

The Hamiltonian $H$ is a special case of a more general Hamiltonian
studied extensively in the literature. It is defined on a domain
$L^2[\mathbb{R}^+ , dx]$ and is of the form
\begin{equation}
H_\alpha = -\frac{d}{dx^2} + \frac{\alpha}{x^2}.
\label{halpha}
\end{equation}
Classically, the system described by this Hamiltonian is scale
invariant ($\alpha$ is a dimensionless constant). However, the quantum
analysis of this operator is much more subtle.  As was shown by
\cite{meetz,narnhofer}, $H_\alpha$ is essentially self- adjoint only
for $\alpha > 3/4$. For $\alpha > 3/4$, the domain of the Hamiltonian
is
\begin{equation}
{\cal D}_0 = \{ \psi \in {\cal L}^2(dx), \psi(0) = \psi'(0) = 0 \}
\end{equation}
For $\alpha \leq 3/4$, this operator is not essentially self-adjoint
(and therefore cannot play the role of a Hamiltonian) and so has to be
``extended'' to another operator. For this case, the deficiency
indices are $\langle 1, 1 \rangle$, and so the self-adjoint extensions
are labeled by a $U(1)$ parameter $e^{iz}$, which labels the domains
${\cal D}_z$ of the Hamiltonian $H_z$. The set ${\cal D}_z$ contains
all the vectors in ${\cal D}_0$, and vectors of the form $\psi_{+} +
e^{iz}\psi_{-}$, where
\begin{eqnarray}
\psi_{+} = x^{1/2}H^{(1)}_{\nu}(x e^{i \pi /4}), \\
\psi_{-} = x^{1/2}H^{(2)}_{\nu}(x e^{-i \pi /4}),
\end{eqnarray}
where $\nu = \sqrt{1/4 + \alpha}$, and $H^{(1,2)}_{\nu}$'s are the
Hankel functions $J_{\nu} \pm i N_{\nu}$. The small $x$ behavior of
$\psi_{+} + e^{iz}\psi_{-}$ is
\begin{eqnarray}  
\psi_{+} + e^{iz}\psi_{-} &\sim& \frac{i x^{1/2}}{\sin(\pi \nu)} \left[
\left( \frac{x}{2} \right)^{\nu} \frac{e^{-3 \pi i \nu /4} - e^{-iz +
3 \pi i \nu /4}}{\Gamma(1+\nu)} \right. \nonumber \\
&&+ \left. \left( \frac{x}{2} \right) ^{-\nu} \frac{e^{iz + i \pi \nu
/4} - e^{-i \pi \nu /4}}{\Gamma(1-\nu)} \right]. \label{smallrbhr}
\end{eqnarray} 

We can now solve the eigenvalue equation for bound states:
\begin{equation}
-\psi'' + \frac{\alpha}{x^2} \psi = -E \psi. 
\label{bseqn}
\end{equation}
For $\alpha \geq 3/4$, there are no bound states. More precisely,
there are no normalizable solutions to the Schr\"odinger equation with
negative energy. However, for $-1/4 \leq \alpha < 3/4$ there is
exactly one bound state of energy $E_b$, where $E_b$ is
\begin{equation}
E_b = E(\nu, z) = \left[ \frac{\sin(z/2 + 3 \pi \nu /4)}{\sin(z/2 +
\pi \nu /4)}\right] ^{1/ \nu},
\end{equation}        
and the corresponding eigenfunction is 
\begin{equation}
\psi = N (\sqrt{E_b}x)^{1/2}[J_{\nu}(i\sqrt{E_b}x) - e^{i \pi
\nu}J_{-\nu}(i\sqrt{E_b}x)]. 
\end{equation}   

The existence of bound states seems to be in contradiction with scale
invariance, since scale invariance implies that there is no length
scale in the problem, whereas the existence of the bound state
provides a scale. This tension can be resolved by looking at how
scaling is implemented in the quantum theory. The scaling operator is
\begin{equation}
\Lambda = \frac{xp+px}{2},
\end{equation}
where $p = -id/dx$. It is easily seen that $\Lambda$ is symmetric on
the domain ${\cal D}$ of $H$, and that for $\alpha > 3/4$,
$\Lambda$ leaves invariant the domain of the Hamiltonian. For $\alpha
\leq 3/4$,
\begin{equation}
\Lambda \psi = x^{3/2}[\psi_{+} + e^{iz}\psi_{-}]'
\end{equation}

The small $x$ behavior of the function $\Lambda \psi$ is of the form
\begin{eqnarray} 
\Lambda \psi &\simeq& \frac{-i \nu x^{1/2}}{\sin \pi \nu} \left[
\left(\frac{x}{2}\right)^{\nu} (2 e^{i \pi /4}-1)\left(
\frac{e^{-3\pi i \nu /4} - e^{i \tilde{z} + 3 \pi i \nu/4}}{\Gamma
(1+\nu)}\right) \right.\nonumber \\  
&+&
\left.\left(\frac{x}{2}\right)^{-\nu} \left(\frac{e^{i \tilde{z} + i \pi
\nu/4} - e^{-i \pi \nu/4}}{\Gamma(1-\nu)}\right) \right] + \cdots,
\end{eqnarray} 
where $\tilde{z} = z +\pi/2$. So $\Lambda \psi$ clearly does not leave
the domain of the Hamiltonian invariant. Scale invariance is thus
anomalously broken, and this breaking occurs precisely when the
Hamiltonian admits non-trivial self-adjoint extensions. This also
explains the quantum mechanical emergence of a length scale, namely
the bound state energy.

We must remark here that there do exist self-adjoint extensions that
preserve scale invariance. For example, if $z = -(\pi \nu /2)$, then
there is no bound state. From the point of view of the domains, the
operator $\Lambda$ leaves this domain invariant, implying that scaling
can be consistently implemented in the quantum theory. 

Now that we know about the subtleties about quantum mechanical
evolution in $1/x^2$ potential, we can apply these ideas to our case.
The potential near the horizon is like $-1/4x^2$ for the problem of
interest.

For the $-1/4x^2$ potential, there are infinite number of bound states
for a given fixed self-adjoint extension $z$. These are given by
\begin{eqnarray} 
\psi_{E_n}(x) &=& N_n \sqrt{x} K_0(\sqrt{E_n}x), \quad n \in
\mathbb{Z}, \\
E_n &=& \exp \left[\frac{\pi}{2} (1-8n) \cot \frac{z}{2} \right], \quad
n \in \mathbb{Z}.
\label{bstates}
\end{eqnarray} 
These are found by solving (\ref{bseqn}) for $\alpha =-1/4$ and
carefully comparing the behavior of the eigenfunctions with the analog
of (\ref{smallrbhr}) which is
\begin{equation}
\psi = e^{-iz/2} (x^{1/2} + i x^{1/2} \ln x) + e^{iz/2} (x^{1/2} - i
x^{1/2} \ln x).
\end{equation}

Returning to the original problem of finding the zero mode solutions,
i.e the solutions to (\ref{conformal}), we see that demanding
self-adjointness of the Hamiltonian gives rise to an infinite number
of bound states labeled by an integer $n$. The zero mode solution is
obtained from (\ref{bstates}) in the $n \rightarrow \infty$ limit. In
particular, the wave function for the solution to (\ref{conformal})
near the horizon is
\begin{equation}
\psi = N_{n} x^{1/2} (1 + \ln (\sqrt{E_{n}} x) ).
\end{equation}
where $E_{n}$ is given by (\ref{bstates}) and $N_{n}$ is an
appropriate normalization factor. Then one takes the limit $n
\rightarrow \infty$. This leads to a solution that is non-zero only at
the horizon, where it peaks, and can be thought of as a `horizon
state'. $E_{n}$ depends on the self-adjointness parameter $z$, which
also corresponds to the boundary condition at the horizon. However, in
the limit $n \rightarrow \infty$, all boundary conditions lead to the
same solution of (\ref{conformal}). Since (\ref{conformal}) is the
time-independent zero angular momentum mode for the scalar field in
{\em all} the aforementioned black hole backgrounds, the above
discussion applies to all those cases. The behavior of the zero mode
found by this method matches that of the numerical zero mode solution
for the Schwarzschild black hole in Fig.1 (where the horizon is at $r
= 50$) apart from minor errors in the numerical interpolaion.

For the one exception, the extremal Reissner-Nordstrom black hole, the
equation (\ref{spham}) is easily solved. The corresponding Hamiltonian
problem for which the solutions to (\ref{spham}) are the zero
eigenvalue solutions was considered in the section above. However, it
does not lead to the kind of non-trivial boundary conditions for the
zero eigenvalue solution as in the other cases. This is because the
self-adjointness analysis of that operator yields only one bound
state. The bound state vanishes for a particular value of the
self-adjointness parameter, as discussed. Therefore, there seems to be
no non-trivial zero mode for the extremal black hole.

\section{Time-independent modes in the plane AdS black hole}
Another black hole solution which can be obtained in the infinite mass
limit from the AdS-Schwarzschild solution, the plane AdS solution, was
discussed in \cite{witten}. The metric for the Euclidean
AdS-Schwarzschild black hole in the infinite mass limit is of the form
\begin{eqnarray} 
ds^2 &=& F(r)d\tau^2 + F^{-1}(r) dr^2 + r^2 \sum_{i~=~1}^{3} dx_{i}^2,
\quad {\rm where} \\ \label{adsbh} 
F(r) &=& (r^2/b^2 - b^2/r^2) 
\end{eqnarray} 
Let us look at a massless scalar field in this background geometry.
One can in principle consider a complex scalar field with charge $q$
and mass $m$, but for simplicity we shall consider only the massless
and uncharged field in the black hole background. The action for such
a field $\Phi$ is
\begin{eqnarray} 
S &=& -\frac{1}{2} \int \sqrt{|g|} g^{ij} (\partial_i \Phi) ( \partial_j
\Phi) d^5 x, \\
&=& -\frac{1}{2} \int_{b}^{\infty}dr     
\int_{0}^{\beta} d\tau \int_{- \infty}^{\infty} d^3 x
\left[ r^3 \left\{ \frac{\dot{\Phi}^2}{F} + F
{(\Phi)'}^2 + 1/r^2 \sum_{i} (\partial_{x_{i}} \Phi)^2 \right\} 
 \right]. \nonumber 
\label{action}
\end{eqnarray}
This action, where the scalar field is the Type IIB dilaton field, has
been discussed in \cite{witten,ooguri,jevicki}. Modes for the field
which are $\tau$ independent are considered, where $\Phi (r,x)~=~f(r)
\exp (ik.x) $. Then the equation of motion for $f(r)$ is
\begin{equation}
- r^{-1} d/dr(r^3 (r^2 - 1/r^2)(df/dr)) + k^2 f = 0,
\label{eqmot}
\end{equation}
where $b=1$ is taken for simplicity. On demanding normalizability of
$f(r)$ w.r.t. the measure $r^3 dr$ and regularity of the solution at
$r=1$, a discrete negative spectrum for $k^2$ was obtained. It was
identified with the glueball spectrum in the boundary theory.

We show below that one can consider (\ref{eqmot}) as an eigenvalue
problem for $k^2$ and examine the operator in this equation for
self-adjointness. As is well known, a self-adjoint operator has only
real eigenvalues, and any wave function in the domain of the operator
can be written in terms of its eigenfunctions. We therefore wish to
find the complete set of $k$ modes such that any function of compact
support in the domain can be expanded in terms of the mode functions.
It is seen that the operator is not self-adjoint, but can be extended
to a self-adjoint operator. However, more general boundary conditions
are required at $r=1$. Then, the spectrum of $k$ is also enlarged to
include a discrete infinity of positive $k^2$ states, and some
negative $k^2$ states as well.

The operator of interest is
\begin{equation}	
T = -r^{-1} d/dr[r^3 (r^2 - 1/r^2) d/dr],
\label{witop}
\end{equation}	
where wave functions are defined on a measure $r^3 dr$.

We can therefore check the operator $T$ for self-adjointness.  We first
check if it is symmetric, i.e. if $(\psi, T\phi)~=~(T^* \psi, \phi)$,
where $\phi~ \epsilon~ D(T)$, and $\psi~\epsilon~D(T^*)$.

If the operator $T$ is symmetric, it is self-adjoint if $(T^* \pm
i)\psi~=~0 $ has no solutions $\psi$ in $D(T^*)$.

But with this measure, we see that the operator is not even
symmetric. We therefore consider the measure $r dr$ which from the
action (\ref{action}) is the natural measure to consider if one is
interested in looking for the eigenvalue problem for the operator
(\ref{witop}). However, this measure is not enough to guarantee
finiteness of the second term in (\ref{action}). Therefore, we take
the domain of functions $D(T)$ to consist of $C_{\infty}$, square
integrable functions with respect to the measure $r dr$ which fall off
at least as $1/r^3$ (or faster than that) that are of compact
support. (Actually, it is enough if they fall of as $1/r^{2 + \delta}$
where $\delta > 0$. For convenience, we take $\delta = 1$, and it does
not affect any of the analysis.)
 
The self-adjointness question is easier to address after a change in
coordinates, following \cite{jevicki}. On making the transformations
\begin{eqnarray}
r^2 &=& \cosh x, \\
A(x) &=& \sqrt{\sinh (2x)} f(x), 
\label{trans}
\end{eqnarray}
the measure becomes $dx/\cosh x$, and (\ref{eqmot}) becomes
\begin{eqnarray}
- 4 \cosh x~ d^2/dx^2 A(x) + 4 \cosh x A(x) - 4 \cosh x A(x)/\sinh
      (2x)^2  \nonumber \\
      = - k^2 A(x) 
\label{jev}
\end{eqnarray}
In these coordinates, the horizon is at $x=0$. Here, one can define
the domain of interest $D(T)$ to consist of $C_{\infty}$, square
integrable functions $A(x)$ with respect to the measure $dx/\cosh x$
and which fall off asymptotically at least as $A(x) \sim
\exp(-3x/2)$. Also, they are of compact support, so
$A(x=0)=A'(x=0)=0$. Then it can be shown that the operator on the
l.h.s of (\ref{jev}) is symmetric, however, the domain of the adjoint
$T^{*}$ is now any normalizable function. Thus, $D(T) \neq
D(T^{*})$. The operator is not self-adjoint. Also, $(T^{*} + i)\psi
=0$ and $(T^{*} - i)\psi = 0$ each have one normalizable solution, as
can be verified numerically (see Fig.2).  If for each eigenvalue $\pm
i$, there is exactly one normalizable solution, then the deficiency
indices of this operator are $(1, 1)$ and it is possible to find
self-adjoint extensions for it. Therefore, one can look for the
self-adjoint extension of this operator. Since a self-adjoint
extension involves only a change of boundary condition at $x = 0$, we
deal with the near-horizon form of (\ref{jev}) for simplicity.

On using the near-horizon ($x$ small) approximation, (\ref{jev}) becomes
\begin{equation}
- (d^2/dx^2) A(x) - \frac{A(x)}{4 x^2} ~~=~~ - \frac{(k^2 +1)A(x)}{4}.
\label{singular}
\end{equation}
This looks like a Hamiltonian problem for a potential $- \frac{1}{4
x^2}$ (which was discussed extensively in the previous section) with
the eigenvalue $- (k^2 + 1)/4$.

The results of the previous section can be applied to the case of
(\ref{singular}) to find the additional states that arise due to the
changed boundary condition. They are given by (\ref{bstates}) with
$E_n~ =~ (k^2 +1)/4$. Thus, there are eigenvalues $k^2$ for each $n$,
and $n$ is any integer. The eigenvalues also depend on the
self-adjoint parameter $z$. There are positive $k^2$ eigenvalues.
There is a possibility of finding some values of $k^2$ with $k^2$
negative too, for which $k^2 < 1$.

What has been done above is a near-horizon analysis of (\ref{jev}). It
is not clear if all of these states are solutions to the complete
equation (\ref{jev}). However, numerically, there seem to exist
normalizable solutions to the complete equation for any positive
$k^2$, provided one also accepts the non-regular solutions that have
not been considered by \cite{witten,ooguri,jevicki}. These are seen to
be irregular {\em only} at the horizon, exactly like the solution in
Fig.2.  Imposing a particular boundary condition at the horizon
demanded by self-adjointness picks out a discrete infinity of
normalizable positive $k^2$ states as above.

A feature of these modes that is immediately noticeable is that they
are irregular at the black hole horizon. However, from considerations
of self-adjointness, they are necessary for expressing any arbitrary,
{\em regular} field configuration in the bulk in terms of a complete
set of mode solutions. In fact, the difference of any two of these
irregular solutions is regular. This is because the irregular
solutions are irregular {\em only} at the horizon, where they behave
as $f_{k} (r) \sim \ln (k (r - 1))$ where $r = 1$ is the horizon.
Taking the difference of two solutions $f_{k1}(r)$ and $f_{k2} (r)$,
we see that the resultant solution is regular at the horizon.
Therefore, any arbitrary regular field configuration in the bulk can
be constructed with regular mode solutions and an even number of
irregular mode solutions.

It may seem that the irregular solutions can be gotten rid of by
shifting the domain of interest a small distance $\epsilon$ away from
the horizon, where $\epsilon > 0$ and repeating the self-adjointness
analysis for this new domain.  However, letting $\epsilon \rightarrow
0$, the irregular solutions reappear. Further, the one parameter
ambiguity in boundary conditions is not resolved.  Letting $\epsilon
\rightarrow 0$ does not pick any particular boundary condition at the
horizon \cite{meetz}.

\section{Discussion}
We find that on examining scalar field theory in the background of the
infinite mass limit of the AdS-Schwarzschild black hole, there are
more time-independent, {\em equilibrium} modes than previously
obtained \cite{ooguri,jevicki}. These are however positive $k^2$
modes. There is a parameter labeling the boundary conditions at the
horizon (for the self-adjoint extension) on which these modes depend.

We analyzed the time-independent, $L = 0$ solutions of the $(3+1)-d$
Schwarzschild and Reissner-Nordstrom black holes, the $(1+1)-d$
dilatonic black hole and the BTZ black hole. There are several
features in these backgrounds that are similar to the case of the
plane AdS-Schwarzschild black hole. In particular, there is again a
one parameter family of boundary conditions labeled by the
self-adjoint parameter $z$ as before. However, now they lead to the
same solution. The solution is a `horizon state', i.e. it is localized
at the horizon. There seems to be no such non-trivial zero mode for
the extremal Reissner-Nordstrom black hole.

Lastly, we would like to speculate on the possible interpretation of
these irregular modes in the boundary theory. As first observed by
\cite{witten}, the modes with negative $k^2$ correspond to glueballs
with mass $k^2$. This correspondence, when applied to the irregular
states, seem to imply the existence of tachyonic glueball states.
Actually, such a scenario is not as exotic as it may appear to be at
first sight. It was pointed out a long time ago by Savvidy
\cite{savvidy}, and also by Nielsen and collaborators \cite{nn,no,ano}
that the perturbative vacuum of $QCD$ is unstable. Considering a
translation invariant background for SU(2) gauge fields, they obtained
the effective one-loop potential. This has the structure of a double
well potential along with an imaginary term signalling the onset of
instability. This persists in SU(N) theories and at finite temperature
\cite{kay}. Our scenario resembles this phenomenon, which seems to be
indicated by the appearance of these modes.

\vskip 1cm

\noindent {\bf Acknowledgements}:
\newline
We would like to thank A. P. Balachandran, S. Kalyana Rama,
R. Parthasarathy, B. Sathiapalan and A. Sen for useful discussions.

\pagebreak

\begin{center}
\epsfxsize 7in
\hspace{0.5cm}
\epsfbox{fig1.epsi}
\end{center}
\vskip 1cm
\begin{tabular}{ll}
\multicolumn{2}{l}
{\bf Figure 1.}\\ 
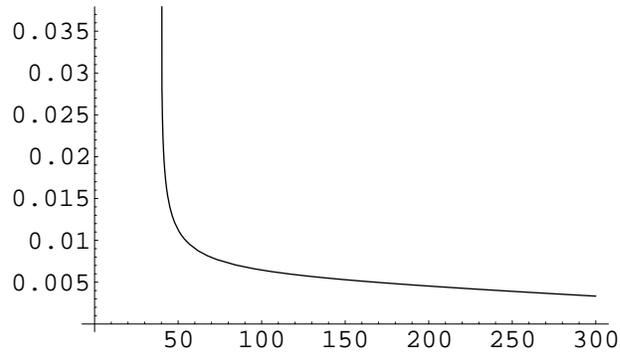
{\sl Absolute value of zero mode solution for a Schwarzschild black}\\ 
{\sl hole with horizon radius $r_{+} = 50$} \\ 
\end{tabular}

\begin{center}
\epsfxsize 7in
\hspace{0.5cm}
\epsfbox{fig2.epsi}
\vskip 1cm
\begin{tabular}{ll}
{\bf Figure 2.} &
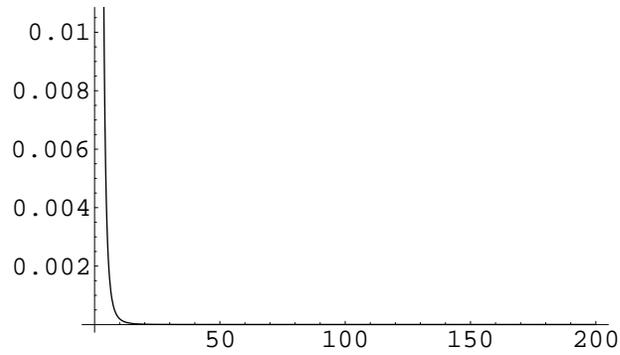
{\sl Absolute value of solution for $k^2 = i$ as a function of $r$}\\
\end{tabular}
\end{center}

\pagebreak
\bibliography{draft4}
\bibliographystyle{unsrt}
\end{document}